\documentclass[12pt]{article}

\usepackage{amsmath}
\usepackage{graphicx}
\usepackage{hyperref}
\usepackage[FIGTOPCAP]{subfigure}

\begin{document}

\title{Why do we use constants of motion while studying the motion of a heavy symmetric top?}

\author{V. Tanr{\i}verdi \\
tanriverdivedat@googlemail.com }
\date{}

\maketitle

\begin{abstract}
	While studying the motion of a heavy symmetric top, in general, constants of motion are used.
	Some students may want to understand the motion in terms of torque, which can lie on their routine based on the usage of Newton's second law.
	However, this is not easy, and examples of this work will illustrate this situation.
	
	In this work, first, we will show the equivalence of torque-angular momentum relation and Euler equations for a heavy symmetric top which give the description of the motion in terms of torque.
	Then, we will study some simple motions of a heavy symmetric top in terms of torque, angular momentum, angular velocities and accelerations, which can help students in understanding rigid body rotations and the necessity of considering the motion of a heavy symmetric top in terms of constants.
	We will also study Perry's historical observational principle on the relation between precession and the rise of the top.

	Keywords: Torque-angular momentum relation, Euler equations, heavy symmetric top
\end{abstract}

\newpage

\section{Introduction}

Classical mechanics books mostly use Lagrangian formalism and constants of motion to explain the motion of a heavy symmetric top \cite{SyngeGriffith, Arya, MatznerShepley, MarionThornton, Goldstein, Taylor77, FowlesCassiday, BargerOlson, JoseSaletan, McCauley, Corinaldesi, KibleBerkshire, Desloge, Symon, Gregory, Greiner, Arnold, LandauLifshitz}.
For an expert, the usage of Lagrangian formalism and constants can be natural. 
On the other hand, students may want further explanations and can seek other explanations.
Trying to use torque to understand the motion can be preferred by students who get used to Newton's second law to understand the motion.
Even some students can wonder that "Do centripetal or Coriolis forces affect the motion of a symmetric top?" though they are not present in the studied reference frames or do not affect the motion.

Euler equations include all effects and describe the rigid body rotation in terms of torque and angular velocities.
But, understanding Euler equations can be difficult for some students due to the presence of extra terms arising from considering the motion in a rotating reference frame.
Therefore, students can encounter different troubles in understanding the motion of a symmetric top.

There are some works giving explanations in terms of torque and related quantities.
Some of these are focused on the basics of the motion and for warming up students \cite{tea1988, cordell2011, kaplan2014}.
An interesting one of these is given by Barker: One of the simplest cases, a model with four point particles resulting in moments of inertia similar to the symmetric top, is considered in this work and the equation describing this simple situation is obtained from the basic principles of rotational motion \cite{barker1960}.
Butikov studied the motion of the heavy symmetric top with simple considerations related to angular momentum and torque without considering angular accelerations \cite{butikov2006}.
Soodak and Tiersten used "resolution principle" which "states that the motion of a rigid body in each short (infinitesimal) time interval may be resolved as a sum or superposition of its torque-free and torque-induced parts." \cite{soodak1994}.
However, we should note that this method can not be used to study the general motion of the heavy symmetric top since governing equations are coupled and torque, in general, can cause angular accelerations in all directions.
Other than these, works of Case and Sch\"{o}nhammer are examples of explanation of the motion in terms of conserved quantities \cite{case1977,shonhammer1998}, and the necessity of these will be seen at the end.

In this work, we will try to help students in understanding and getting over mentioned troubles.
In section \ref{one}, we will explicitly show the equivalence of Euler equations with torque-angular momentum relation, which is not shown in classical mechanics books.
This can help students understanding the extra terms in Euler equations.

In section \ref{two}, we will study some simple cases in terms of angular accelerations, angular velocities, torque and angular momentum.
Some books focused on rigid body rotations or the motion of a symmetric top give similar explanations \cite{Routh, Crabtree, ArnoldMaunder, Scarborough, Gray}. 
Different from these works, we will focus more on the regular precession and the effect of precession angular velocity. 
We will also consider the motion with cusps which is also considered by Gray with some differences \cite{Gray}.
We will explain simple motions in terms of angular acceleration and angular velocity which can help students to realize the motion of a heavy symmetric top.
We will also give explanations on the effect of torque, and it will be clear at the end that torque alone is not enough to explain the motion of a heavy symmetric top.
It will be also clear that the usage of angular accelerations and angular velocities is not enough to understand the whole motion, which gives a clue to the necessity of the usage of constants.

The results of section \ref{twoc} will explain the relationship between the rise of the top and precessional angular velocity which is related to the observational principle: "Hurry on the precession, and the body rises in opposition to gravity." 
This principle is firstly given by Perry \cite{Perry}, and later given by Crabtree \cite{Crabtree}.
We will conclude and explain that Perry's observational principle is conditionally valid in the conclusion part, which is not explained previously.

\section{Equivalence of Euler equations with torque-angular momentum relation}
\label{one}

The relation between torque and angular momentum is given by \cite{Goldstein}
\begin{equation}
	\vec \tau= \frac{d \vec L} {d t},
	\label{tamr}
\end{equation}
where $\vec \tau$ is torque, and $\vec L$ is the angular momentum. 
This equation tells us that "The rate of change of angular momentum is equal to torque." 
This can be considered obvious, but examples in the next section will show that some results of this equation are not that obvious.

The angular momentum can be written as a tensor product
\begin{equation}
	\vec L = I \vec w,
	\label{amom}
\end{equation}
where $I$ is the moment inertia tensor, and $\vec w$ is the angular velocity.
Moment of inertia tensor, in general, can be written as
\begin{equation}
        I=
        \begin{bmatrix}
        I_{x'x'} & I_{x'y'}  &  I_{x'z'} \\
        I_{y'x'} & I_{y'y'}  &  I_{y'z'} \\
        I_{z'x'} & I_{z'y'}  &  I_{z'z'} 
        \end{bmatrix},
\end{equation}
where $I_{i'i'}$'s are moments of inertia, and $I_{i'j'}$'s are products of inertia.
Any component of inertia tensor can be found by using \cite{Goldstein}
\begin{equation}
	I_{i'j'}=\int_V \rho(\vec r ) (r^2 \delta_{ij}-r_i r_j) dV.
	\label{momg}
\end{equation}
Calculating components of the moment of inertia tensor in this way can be tedious in the stationary reference frame. 
Instead of this method, one can calculate components of the moment of inertia tensor in a coordinate system whose axes are principal axes of the rigid body and then transform it.
In such a coordinate system, moments of inertia are different from zero and products of inertia are equal to zero, and there is always such a coordinate system \cite{Goldstein}.
The mass distribution of the rigid body with respect to a principal axis is responsible for this and it is such that the integral given by equation \eqref{momg} is equal to zero when $i' \ne j'$. 
For a better understanding of moments of inertia and products of inertia, one should study Euler's derivation \cite{Euler1752}.

In this work, this coordinate system or a coordinate system whose axes are parallel to the mentioned coordinate system will be named as body reference frame and moments of inertia will be shown by $I_i$ instead of $I_{ii}$ in the body reference frame as usual. 
And then, moments of inertia tensor in the body reference frame can be written as
\begin{equation}
        I_{b}=
        \begin{bmatrix}
           I_x & 0 & 0 \\
           0 & I_y & 0\\
           0 & 0 & I_z
        \end{bmatrix} .
	\label{moib}
\end{equation}

\begin{figure} [h!]
        \begin{center}
          \includegraphics[width=5.7cm]{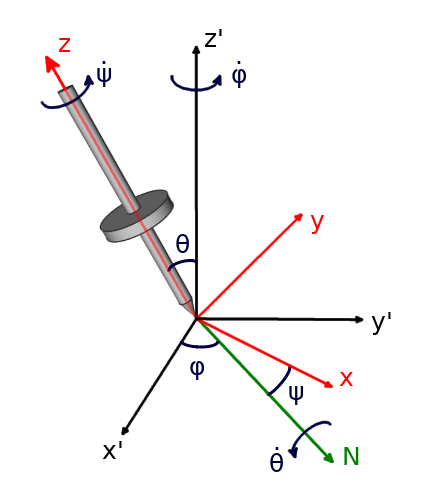}
          \caption{A symmetric top, body reference frame $S(x,y,z)$, stationary reference frame $S'(x',y',z')$, line of nodes $N$, Euler angles and angular velocities in terms of Euler angles.
                $\theta$ is the angle between $z'$-axis and $z$-axis, $\phi$ is the angle between $x'$-axis and the line of nodes, and $\psi$ is the angle between the line of nodes and $x$-axis.}
         \label{fig:eag}
        \end{center}
\end{figure}

After finding the moment of inertia tensor in the body reference frame, 
one can find it in any reference frame by using a transformation matrix.
The transformation matrix from the body reference frame to the stationary reference frame in terms of Euler angles can be obtained as \cite{Goldstein} 
\footnotesize
\begin{equation}
	S=
  	\begin{bmatrix}
            \cos \phi \cos \psi - \cos \theta \sin \phi \sin \psi & -\cos \phi \sin \psi -\cos \theta \sin \phi \cos \psi & \sin \theta \sin \phi \\
            \sin \phi \cos \psi + \cos \theta \cos \phi \sin \psi & -\sin \phi \sin \psi +\cos \theta \cos \phi \cos \psi & -\sin \theta \cos \phi \\
            \sin \theta \sin \psi & \sin \theta \cos \psi & \cos \theta
         \end{bmatrix}.
 \label{eqtm}
\end{equation}
\normalsize
The inverse of $S$ can be found by using $S^{-1}_{ij}=S_{ji}$.
The Euler angles together with a symmetric top, body reference frame, stationary reference frame, line of nodes and angular velocities can be seen in figure \ref{fig:eag}.

For a symmetric top, $I_x=I_y$, the components of moments of inertia tensor in the stationary reference frame can be obtained by using $I=S I_b S^{-1}$ \cite{Goldstein} as
\begin{eqnarray}
        I_{x'x'}&=& I_x \cos^2 \phi   +I_x \cos^2 \theta \sin^2 \phi +I_z \sin^2 \theta \sin^2 \phi, \nonumber \\
        I_{y'y'}&=& I_x \sin^2 \phi   +I_x \cos^2 \theta \cos^2 \phi +I_z \sin^2 \theta \cos^2 \phi, \nonumber \\
        I_{z'z'}&=& I_x \sin^2 \theta +I_z \cos^2 \theta,  \label{moi} \\ 
        I_{x'y'}&=& I_{y'x'}= (I_x-I_z ) \sin^2 \theta \sin \phi \cos \phi, \nonumber \\
        I_{x'z'}&=& I_{z'x'}= (I_z-I_x ) \sin \theta \cos \theta \sin \phi,  \nonumber \\
        I_{y'z'}&=& I_{z'y'}= (I_x-I_z) \sin \theta \cos \theta \cos \phi. \nonumber
\end{eqnarray}

Then, by using equation \eqref{amom}, one can write the angular momentum in the stationary reference frame as
\begin{eqnarray}
        \vec L&=& (I_{x'x'} w_{x'}+I_{x'y'} w_{y'}+I_{x'z'} w_{z'}) \hat x' +(I_{y'x'} w_{x'}+I_{y'y'} w_{y'}+I_{y'z'} w_{z'}) \hat y' \nonumber \\
              & & +(I_{z'x'} w_{x'}+I_{z'y'} w_{y'}+I_{z'z'} w_{z'}) \hat z',
              \label{angmoms}
\end{eqnarray}
where angular velocities in the stationary reference frame in terms of Euler angles can be written as
\begin{eqnarray}
        w_{x'}&=& \dot \theta \cos \phi+ \dot \psi \sin \theta \sin \phi, \nonumber \\
        w_{y'}&=& \dot \theta \sin \phi- \dot \psi \sin \theta \cos \phi, \label{angvelsrf}\\
        w_{z'}&=& \dot \phi+ \dot \psi \cos \theta, \nonumber
\end{eqnarray}
where $\dot \theta$ is the nutation angular velocity,
$\dot \phi$ is the precession angular velocity, 
and $\dot \psi$ is the spin angular velocity, 
and they define rotation around the line of nodes, stationary $z'$-axis and body $z$-axis, respectively.
The gravitational torque for the heavy symmetric top in the stationary reference frame can be written as
\begin{equation}
        \vec \tau_g=Mgl \sin \theta (\cos \phi \hat x'+\sin \phi \hat y').
        \label{tgrs}
\end{equation}
From this relation, it can be seen that the gravitational torque is in the direction of the line of nodes.
If an object having zero angular momentum is subject to this torque alone, the angle $\theta$ will increase, however, the motion of a spinning symmetric top is much more complicated due to the presence of the angular momentum.
There will be some examples to this complicated situation in the next section.

Since we have obtained torque and angular momentum in the stationary reference frame, we can use torque-angular momentum relation to find the angular accelerations.
To do this, we should take the time derivative of the angular momentum.
Equations \eqref{moi} and \eqref{angmoms} show that while finding the time derivative of angular momentum, we need to take into account the changes of moments of inertia and products of inertia.
This can easily be understood by considering the motion of the heavy symmetric top:
As the top rotates, angles $\theta$ and $\phi$ change, then distances to the axes of the stationary reference frame change, and since moments of inertia and products of inertia depend on the distances,
they change at each instant, and then one should also take into account their time derivatives while finding the time derivative of the angular momentum.
After finding the time derivative of the angular momentum, from the torque-angular momentum relation, one can obtain
\footnotesize
\begin{eqnarray}
        Mgl \sin \theta \cos \phi &=& I_x \ddot \theta \cos \phi + (I_z-I_x) \ddot \phi \sin \theta \cos \theta \sin \phi + I_z \ddot \psi \sin \theta \sin \phi \nonumber \\
                                  & & - 2 I_x \dot \theta \dot \phi \cos^2 \theta \sin \phi + I_z \dot \theta \dot \phi \sin \phi (\cos^2 \theta -\sin ^2 \theta) + I_z \dot \theta \dot \psi \cos \theta \sin \phi \nonumber \\
                                  & & + I_z \dot \phi \dot \psi \sin \theta \cos \phi+ (I_z-I_x) \dot \phi^2 \sin \theta \cos \theta \cos \phi, \nonumber \\
        Mgl \sin \theta \sin \phi &=& I_x \ddot \theta \sin \phi - (I_z-I_x) \ddot \phi \sin \theta \cos \theta \cos \phi - I_z \ddot \psi \sin \theta \cos \phi  \label{deqnssrf} \\
                                  & & + 2 I_x \dot \theta \dot \phi \cos^2 \theta \cos \phi - I_z \dot \theta \dot \phi \cos \phi (\cos^2 \theta -\sin ^2 \theta) - I_z \dot \theta \dot \psi \cos \theta \cos \phi \nonumber \\
                                  & & + I_z \dot \phi \dot \psi \sin \theta \sin \phi+ (I_z-I_x) \dot \phi^2 \sin \theta \cos \theta \sin \phi, \nonumber \\
                                0 &=& I_z \ddot \psi \cos \theta + \ddot \phi (I_x \sin^2 \theta+ I_z \cos^2 \theta)-I_z \dot \theta \dot \psi \sin \theta \nonumber \\
				  & & - 2 (I_z-I_x) \dot \theta \dot \phi \sin \theta \cos \theta. \nonumber
\end{eqnarray}
\normalsize
By using these three equations, after some algebra, one can obtain angular accelerations in terms of Euler angles for a spinning heavy symmetric top as
\begin{eqnarray}
        \ddot \theta &=& \frac{\sin \theta}{I_x} \left[ (I_x-I_z)\dot \phi^2 \cos \theta-I_z \dot \phi \dot \psi + Mgl \right], \label{diffeqnstheta} \\
	\ddot \phi   &=& \frac{\dot \theta}{I_x \sin \theta}\left[ I_z \dot \psi +(I_z-2 I_x) \dot \phi \cos \theta \right],  \label{diffeqnsphi} \\
	\ddot \psi   &=& \frac{\dot \theta}{I_x} \left[ -\cot \theta \left( I_z \dot \psi +(I_z-2 I_x) \dot \phi \cos \theta \right)+ I_x \dot \phi \sin \theta \right]. \label{diffeqnspsi}
\end{eqnarray}
As it is seen, finding angular accelerations in the stationary reference frame is a bit cumbersome.

After seeing the change of moments of inertia tensor in the stationary reference frame \cite{Euler1752}, Euler has found a simpler way to study rigid body rotations: Studying in the body reference frame whose axes are principal axes and fixed to the body \cite{Euler1765}.
We have already given some explanations on this reference frame.
This simpler way results in Euler equations for rigid body rotations.
These equations can be obtained by writing the torque-angular momentum relation in the body reference frame which is a rotating reference frame,
and to do this, one can write the time derivative of angular momentum in a rotating reference frame by using $(d \vec L /d t)_s=(d \vec L /d t)_b+\vec w \times \vec L$ \cite{Goldstein} where subscripts $s$ and $b$ indicate the stationary and body reference frames, respectively.
We should note that Euler used direction cosines to obtain these equations.

Now, we will use Euler equations to find angular accelerations.
For a symmetric top, Euler equations can be obtained by using $(d \vec L /d t)_b+\vec w \times \vec L$ as \cite{Goldstein}
\begin{eqnarray}
          \tau_x&=&I_x \dot w_x - w_y w_z(I_x-I_z),  \nonumber \\
          \tau_y&=&I_y \dot w_y - w_z w_x(I_z-I_x),  \label{eest} \\
          \tau_z&=&I_z \dot w_z ,  \nonumber
\end{eqnarray}
where $\tau_i$'s are components of torque and $w_i$'s are components of angular velocity in the body reference frame.
The components of the angular velocity in the body reference frame can be written in terms of Euler angles as
\begin{eqnarray}
          w_{x}&=& \dot \phi \sin \theta \sin \psi + \dot \theta \cos \psi, \nonumber \\
          w_{y}&=& \dot \phi  \sin \theta \cos \psi - \dot \theta \sin \psi, \label{angvel}\\
          w_{z}&=& \dot \phi \cos \theta + \dot \psi.  \nonumber
\end{eqnarray}
For a heavy symmetric top, the gravitational torque, given also in equation \eqref{tgrs}, can be written in the body reference frame as
\begin{equation}
        \vec \tau_g=Mgl \sin \theta (\cos \psi \hat x-\sin \psi \hat y).
        \label{tgr}
\end{equation}

Then, by using equations \eqref{angvel} and \eqref{tgr} in Euler equations, after some algebra, one can obtain the following equations 
\begin{align*}
	\ddot \theta &= \frac{\sin \theta}{I_x} \left[ (I_x-I_z)\dot \phi^2 \cos \theta-I_z \dot \phi \dot \psi + Mgl \right], \tag{\ref{diffeqnstheta}} \\
	\ddot \phi   &= \frac{\dot \theta}{I_x \sin \theta}\left[ I_z \dot \psi +(I_z-2 I_x) \dot \phi \cos \theta \right],  \tag{\ref{diffeqnsphi}} \\
	\ddot \psi   &= \frac{\dot \theta}{I_x} \left[ -\cot \theta \left( I_z \dot \psi +(I_z-2 I_x) \dot \phi \cos \theta \right)+ I_x \dot \phi \sin \theta \right], \tag{\ref{diffeqnspsi}}
\end{align*}
which are the same as the equations obtained from the torque-angular momentum relation.
One can see that the usage of Euler equations are much easier than the usage of toque-angular momentum relation.

This equivalence explicitly shows that Euler equations are not different from the torque-angular momentum relation as it should be.
The extra terms in Euler equations coming from $\vec w \times \vec L$ are the results of considering the motion in an accelerated reference frame,
and they are inertial terms and can be considered as inertial torque.
Inertial torque is the analog of inertial forces or fictitious forces which are present in the equations of motion when calculations are done in an accelerated reference frame. 
The same equations can also be obtained from Lagrangian formalism \cite{Tanriverdi_cagrisd}, which shows that there is not any difference between different approaches as it should be.

\section{Motion of the top}
\label{two}

In this part, we will consider a few different motions of a heavy symmetric top in terms of angular accelerations, angular velocities, torque and angular momentum, and $Mgl$ is considered as always positive.
We will mainly focus on the effect of precession angular velocity to the motion.
The numerical solutions are obtained by numerically integrating angular accelerations \cite{Tanriverdi_cagrisd}.

\subsection{Regular precession}
\label{twoa}

Now, we will consider one of the simplest motions of a spinning heavy symmetric top: Regular precession.
During the regular precession, the top precesses regularly, and $\dot \phi$, $\dot \psi$ and $\theta$ are constants.
Then, one can say that in the regular precession, $\ddot \theta$ is always equal to zero, and initial values should ensure it.
From equations \eqref{diffeqnsphi} and \eqref{diffeqnspsi}, since $\dot \theta$ is always equal to zero, one can also say that $\ddot \phi$ and $\ddot \psi$ are always zero as it should be.

For the regular precession, one can ask "Why angular velocities do not change in spite of the presence of gravitational torque?"
The answer to this question is "Torque is balanced out by the change in the direction of the angular momentum."

Let us analyze this in detail in terms of angular accelerations.
If we look at equations \eqref{diffeqnstheta}, \eqref{diffeqnsphi} and \eqref{diffeqnspsi}, we see that the effect of the gravitational force is only seen in the angular acceleration $\ddot \theta$.
We have already mentioned that torque is in the direction of the line of nodes, and it results in an increase of $\theta$ when the top does not spin.
On the other hand, in the regular precession, there is not any change in $\theta$ though the same torque is present.
We can see why this torque does not change $\theta$ by using equation \eqref{diffeqnstheta}.
To get the regular precession, $\ddot \theta$ should be equal to zero, and $\dot \phi$ should make the right-hand side of equation \eqref{diffeqnstheta} equal to zero 
\begin{equation}
	(I_x-I_z)\dot \phi^2 \cos \theta-I_z \dot \phi \dot \psi+Mgl=0.
	                \label{quadr}
\end{equation}
From this equation, one can say that torque that shows itself in this equation as $Mgl$ after simplification of $\sin \theta$ can be balanced out by the precession angular velocity which is what happens at the regular precession.

Now, let us consider torque and angular momentum for the regular precession.
The magnitude of torque does not change since $\theta$ is constant, and its direction is always in the direction of the line of nodes which rotates around stationary $z'$-axis with $\dot \phi$ and perpendicular to the angular momentum in this case.
$\theta$ does not change and speeds of the rotations around the stationary $z'$-axis and body $z$-axis do not change, and accordingly, the angular momenta in $\hat z'$ direction and $\hat z$ direction are constant.
Since $\dot \phi \ne 0$, the direction of angular momentum always changes and rotates around the stationary $z'$-axis in the regular precession.
Then by considering the torque-angular momentum relation, one can say that to get regular precession, the value of $\dot \phi$ should be finely tuned to give just the right amount of angular momentum change at each time interval which must be exactly equal to torque.

One can find the angular momentum $\vec L (t+dt)$ by adding $\vec \tau dt$ to the angular momentum vector $\vec L (t)$ which can be seen from $\vec \tau= d\vec L / dt$.
Some people may think that there should be a single value of $\vec L (t)$ for the regular precession at a specific angle $\theta$ since $\vec \tau$ is single-valued at that angle.
However, this consideration does not completely fit in the situation: There are, in general, two solutions for the regular precession at that specific angle $\theta$ with a definite $\dot \psi$ which can be seen from equation \eqref{quadr}.
This situation shows that $\vec \tau= d\vec L / dt$ means that the rate of change of angular momentum is equal to torque and this rate of change can be seen at more than one configuration for a specific torque.

\begin{figure}[h]
        \begin{center}
        \includegraphics[width=5.5cm]{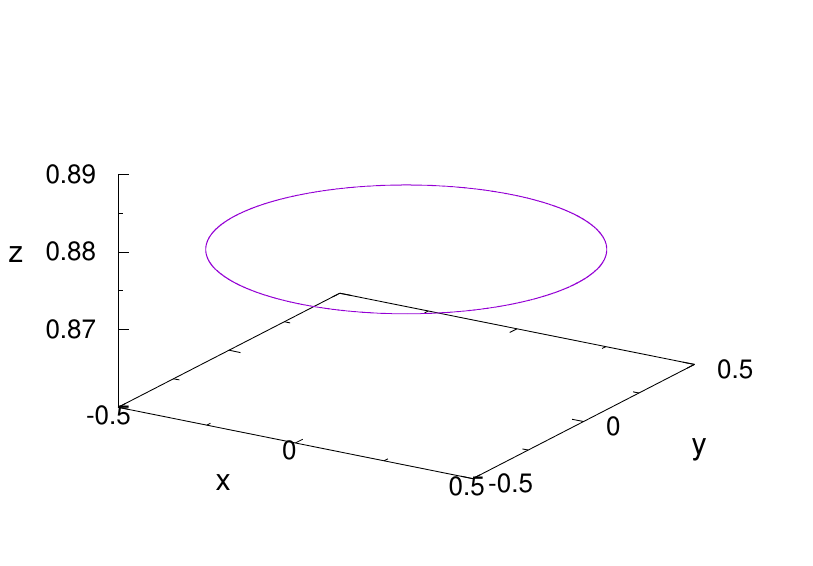}
		\caption{Shapes for the locus on the unit sphere for the regular precession. Initial values and constants: $\theta_0=0.5 \,rad$, $\phi_0=0$, $\psi_0=0$, $\dot \theta_0=0 $, $\dot \phi_0=6.50 \,rad\,s^{-1} $, $\dot \psi_0=200 \,rad\,s^{-1}$, $I_x=0.000228 \,kg \,m^2$, $I_z=0.0000572 \,kg \,m^2$ and $Mgl=0.068J$.
		The animated version can be found at \href{https://youtu.be/EuMSPGxA2Sc}{https://youtu.be/EuMSPGxA2Sc}.
                }
        \label{fig:uefftt_1}
        \end{center}
\end{figure}

In figure \ref{fig:uefftt_1}, we see an example of the regular precession: A spinning heavy symmetric top has an initial precession angular velocity together with zero nutation angular velocity, 
and that precession angular velocity results in the precession of the top regularly with constant $\theta$ without any change in $\dot \phi$ and $\dot \psi$. 

\subsection{Motion with cusps}
\label{twob}

\begin{figure}[h]
        \begin{center}
                \subfigure[$\theta$]{
                \includegraphics[width=3.0cm]{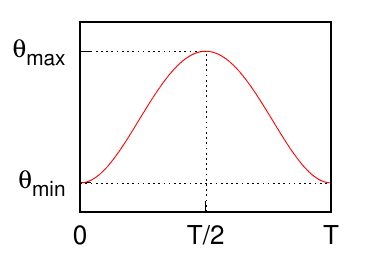}
                }
                \subfigure[$\dot \theta$]{
                \includegraphics[width=3.0cm]{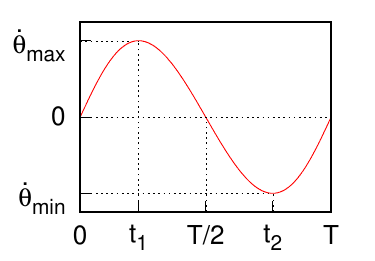}
                }
                \subfigure[$\dot \phi$]{
                \includegraphics[width=3.0cm]{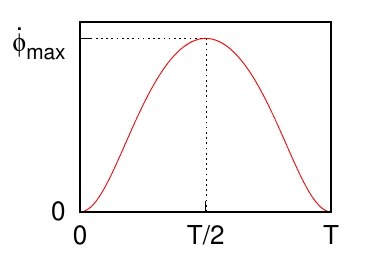}
                }
                \subfigure[$\dot \psi$]{
                \includegraphics[width=3.0cm]{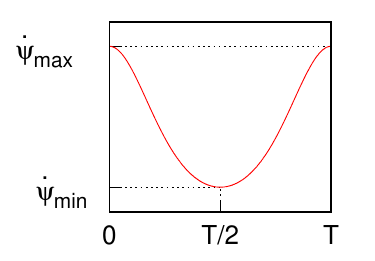}
                }
                \caption{(a) $\theta$, (b) $\dot \theta$, (c) $\dot \phi$ and (d) $\dot \psi$ for motion with cusps.
	                }
               \label{fig:ttdfd_2}
        \end{center}
\end{figure}

Now, we will consider what happens to a spinning heavy symmetric top ($\dot \psi>0$) if there is not any precession angular velocity and nutation angular velocity at $t=0$.
One can see changes in $\theta$ and angular velocities in figure \ref{fig:ttdfd_2} for such a case.

Let us analyze these changes by using angular accelerations.
From equations \eqref{diffeqnsphi} and \eqref{diffeqnspsi}, one can say that angular accelerations $\ddot \phi$ and $\ddot \psi$ are equal to zero at $t=0$ since $\dot \theta=0$ at that instant.
From equation \eqref{diffeqnstheta}, one can easily say that there will be an angular acceleration related to $\theta$ at $t=0$ resulting with a positive $\dot \theta$ since $\dot \phi(t=0)=0$.
This shows that the top will fall at the beginning, but this fall is only the beginning of the motion.
Therefore, just after $t=0$, $\theta$ will change and $\phi$ will not change.
After the initial change in $\theta$, one can not comment on the motion with simple statements, and one needs to speak on the motion dynamically since each change causes a sequence of changes.
Since $\dot \theta$ is not equal to zero anymore, there will be changes in $\dot \phi$ and $\dot \psi$.
After the start of the motion, one can say from equation \eqref{diffeqnsphi} that $\dot \phi$ will increase since $\dot \theta>0$ and $\dot \psi>0$.
At the beginning, $\dot \phi$ is small and $Mgl$ term is dominant at right-hand side of equation \eqref{diffeqnstheta} and $\theta$ continues to increase.
And, $\dot \phi$ will increase as $\theta$ increases in this case since $\ddot \phi > 0$ as $\dot \theta>0$ and $\dot \psi>0$.
And after some time, at $t=t_1$, $\dot \phi$ becomes big enough to make the right-hand side of equation \eqref{diffeqnstheta} equal to zero, i.e. $\ddot \theta=0$, similar to the regular precession.
However, at that moment, $\dot \theta$ is not equal to zero (in fact it is at the maximum), and one can not observe the regular precession, and $\theta$ continues to increase.
Some more time is required to make $\dot \theta$ equal to zero, and $\theta$ continues to increase with a negative $\ddot \theta$ till $t=T/2$.
At $t=T/2$, $\dot \theta$ becomes equal to zero, and $\theta$ and $\dot \phi$ reach their maximum.
From $t=0$ to $t=T/2$, $\dot \psi$ decreases when $\theta$ is small enough as shown in figure \ref{fig:ttdfd_2}(d).
If $\theta$ is not small enough, $\dot \psi$ can decrease at some part of the motion and can increase at the other part or can only increase from $t=0$ to $t=T/2$ \cite{Tanriverdi_abdffrnt}.
At the bottom point, $\theta=\theta_{max}$ or $t=T/2$, the top starts to rise since $\ddot \theta$ is negative and $\dot \theta=0$, and the mentioned changes take place in the reverse order till $\theta$ reaches its initial value and $\dot \phi$ becomes zero. 
This procedure, from $t=0$ to $t=T$, repeats itself periodically.
We should note that, in general, $t_1 \ne T/4$ and $t_2 \ne 3T/4$.

For an ordinary symmetric top, spin angular velocity can initially be positive or negative.
In the mentioned example, $\dot \psi$ is considered as positive; and if it were negative, the overall precession would be in the reverse direction.
This can be seen from equation \eqref{diffeqnsphi}; if $\dot \psi$ is positive (negative), then $\ddot \phi$ becomes positive (negative). 

\begin{figure}[h!]
        \begin{center}
               \includegraphics[width=5.5cm]{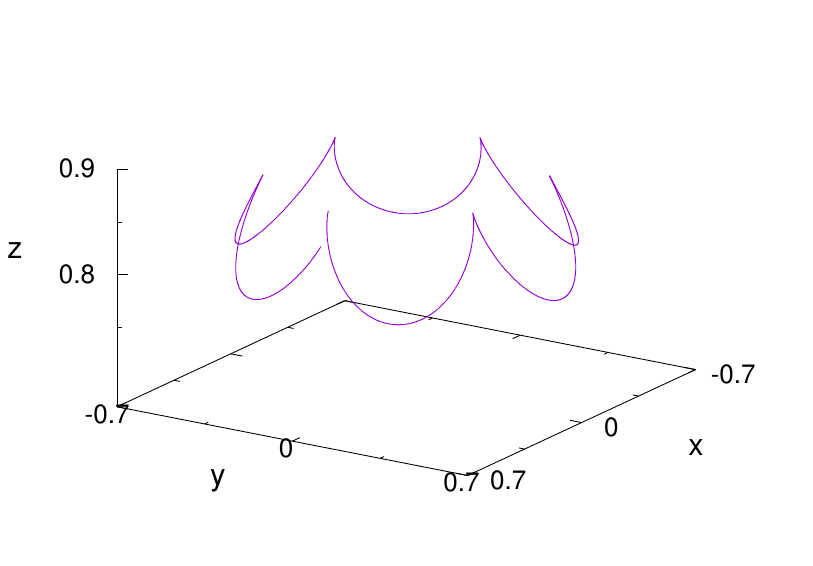}
               \caption{Shapes for the locus on the unit sphere for motion with cusps. 
		Initial values and constants: $\theta_0=0.5 \,rad$, $\phi_0=0$, $\psi_0=0$, $\dot \theta_0=0 $, $\dot \phi_0=0 $, $\dot \psi_0=200 \,rad\,s^{-1}$, $I_x=0.000228 \,kg \,m^2$, $I_z=0.0000572 \,kg \,m^2$, $Mgl=0.068J$.
		The animated version can be found at \href{https://youtu.be/zTcVg25xF44}{https://youtu.be/zTcVg25xF44}.
                }
       \label{fig:uefftt_3}
       \end{center}
\end{figure}

In figure \ref{fig:uefftt_3}, one can find a three-dimensional plot for an example of the mentioned situation, and one can see that there are cusps at the motion.
The number of cusps can change for different initial values.

\begin{figure}[h!]
	\begin{center}
		\subfigure[]{
		\includegraphics[width=3.9cm]{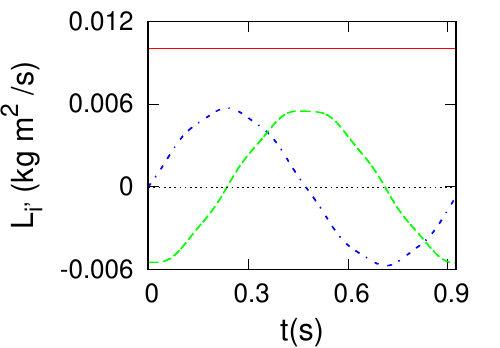}
	       }
		\subfigure[]{
		\includegraphics[width=3.5cm,height=3.0cm]{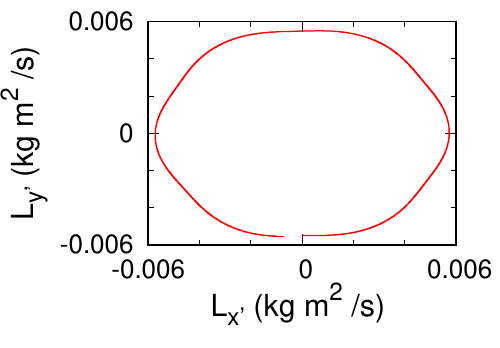}
		}
	      \subfigure[]{
	      \includegraphics[width=3.9cm]{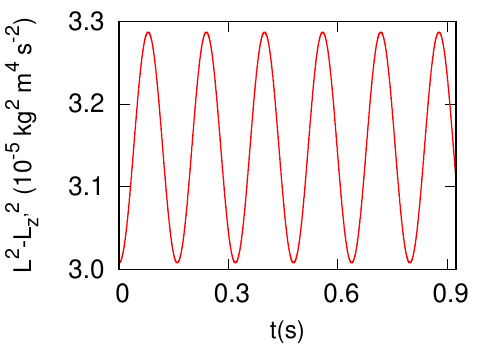}
	      }       
		\caption{(a) Change of $L_{x'}$ (dotted-dashed blue curve), $L_{y'}$ (dashed green curve), $L_{z'}$ (continuous red line) for the motion given in figure \ref{fig:uefftt_1}.
	        (b) Projection of $\vec L$ onto $x'y'$-plane,
		(c) Change of $L^2-L_{z'}^2$.
       }
	\label{fig:lxlylz}
	\end{center}
\end{figure}

Motions with $\dot \theta \ne 0$ is very hard to explain verbally in terms of torque and change of angular momentum.
It is not simple to find $\vec L (t+dt)$ by vectorially adding $\vec \tau dt$ to $\vec L(t)$ since there are angular accelerations.
One can see this by considering the up-down motion: $\theta$ increases at some part of the motion and decreases at the other parts though torque is always in the direction of the line of nodes which results in an increase of $\theta$ if there is not any angular momentum at the beginning. 

Here, we will just show the changes of angular momentum in figure \ref{fig:lxlylz} and mainly describe the changes for the considered case in figure \ref{fig:uefftt_3}.
In figure \ref{fig:lxlylz}(a), one can see the changes of $L_{x'}$, $L_{y'}$ and $L_{z'}$ components of angular momentum which can be found by using equation \eqref{angmoms}.
It can be seen that $L_{z'}$ is constant, and $L_{x'}$ and $L_{y'}$ are changing with time, and their change is not simple sinusoidal.
In figure \ref{fig:lxlylz}(b), the projection of angular momentum into $x'y'$-plane can be seen, 
and it can be seen that there are some oscillations from the circle.
Though the motion covers the change of $\phi$ from $0$ to $2 \pi$, the projection of angular momentum is not closed due to oscillations of $L_{x'}$ and $L_{y'}$.
In figure \ref{fig:lxlylz}(c), one can see the change of $L^2-L_{z'}^2=L_{x'}^2+L_{y'}^2$.
Since $L_{z'}$ is constant, changes of $L^2$ are the same as changes of $L^2-L_{z'}^2$ with a constant difference.
It can be seen that the change of $L^2-L_{z'}^2$ is similar to simple sinusoidal, and the number of oscillations is the same as the number of nutations.
One can see that these changes are not simple, and it is very hard to understand and explain the motion of the top's symmetry axis by considering changes of angular momentum shown in figure \ref{fig:lxlylz}.
On the other hand, one can see a relation in these changes in terms of energy: One can see that at the beginning, $L^2-L_{z'}^2$ is minimum, where $\theta$ is minimum, and then one can say that as the top falls, the decrease in its potential energy shows itself as an increase in the angular momentum or kinetic energy.

In this motion, in the beginning, torque causes a change both in the magnitude and direction of the angular momentum, which results in a sequence of many changes.
All of these changes obey torque-angular momentum relation $\vec \tau=d \vec L/ d t$ since they are obtained from the resultant equations, i.e. equations \eqref{diffeqnstheta}, \eqref{diffeqnsphi} and \eqref{diffeqnspsi}.
This shows that at each moment the rate of change of angular momentum is equal to torque.
However, one can not understand the motion by considering only torque.
We should note that the magnitude of torque changes as $\theta$ changes and its direction is always in the direction of the line of nodes that rotates with $\dot \phi$ around the stationary $z'$-axis.

\subsection{Motion for different precession angular velocities}
\label{twoc}

\begin{figure}[h!] 
        \begin{center}
              \includegraphics[width=5.50cm]{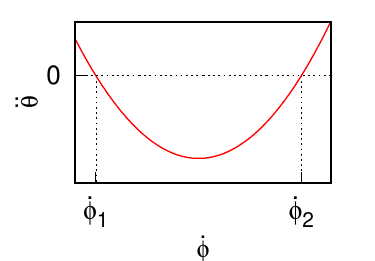}
		\caption{$\ddot \theta$ as a function of $\dot \phi$ when $\theta$ and $\dot \psi$ are held constant when $\dot \psi_0>0$ and $\theta_0<\pi/2$ for a top safisfying $I_x>I_z$.
                }
        \label{fig:ddot_theta}
        \end{center}
\end{figure}

As a next step, we will consider what will happen for different values of $\dot \phi_0 $ when $\dot \theta=0$ and $\dot \psi \ne 0$ at $t=0$.
Before giving examples, we should consider equation \eqref{diffeqnstheta} in a bit more detail.
The right-hand side of equation \eqref{diffeqnstheta} is second degree in $\dot \phi$, and its graph with respect to $\dot \phi$ opens upward when $(I_x-I_z) \cos \theta>0$ and downward when $(I_x-I_z) \cos \theta<0$. 
One can see an example of it opening upward in figure \ref{fig:ddot_theta}.
We used equation \eqref{diffeqnstheta} to write down equation \eqref{quadr}, and its roots are values of $\dot \phi$ making $\ddot \theta$ equal to zero.
These values can be obtained from the solution of quadratic equation as
\begin{equation}
	      \dot \phi_{1,2} =\frac{I_z \dot \psi \pm \sqrt{I_z^2 \dot \psi^2 -4 (I_x-I_z)Mgl \cos \theta}}{2 (I_x-I_z) \cos \theta}.
\end{equation} 
For a positive discriminant and given $\theta$ and $\dot \psi$ values, these two roots can be found. 
If $(I_x-I_z) \cos \theta>0$, both roots have the same sign; and if $(I_x-I_z) \cos \theta<0$, roots have differents signs.
When both roots have the same sign, their sign is the same as the sign of $\dot \psi$.

Now, we will consider what will happen for different values of $\dot \phi_0>0$ when $\dot \psi_0>0$, $\dot \theta=0$ and $\theta_0<\pi/2$ for a top safisfying $I_x>I_z$.
With this conditions, the resultant graph of $\ddot \theta$ is similar to figure \ref{fig:ddot_theta}.
One can say that if $\dot \phi_0$ is equal to either $\dot \phi_1$ or $\dot \phi_2$, then the regular precession is observed; if $\dot \phi_0$ is between $\dot \phi_1$ and $\dot \phi_2$, then the top rises; and if $\dot \phi_0$ is smaller than $\dot \phi_1$ or greater than $\dot \phi_2$, then the top falls.

\begin{figure}[h!]
        \begin{center}
		\subfigure[]{\includegraphics[width=4.70cm]{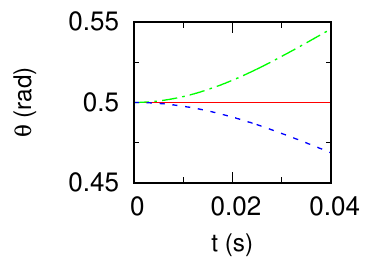}}
		\subfigure[]{\includegraphics[width=4.70cm]{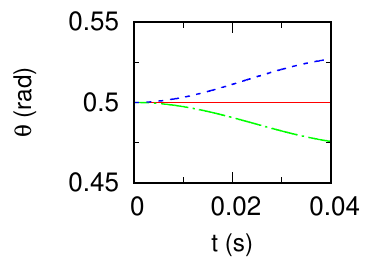}}	
		\caption{Changes of $\theta$ for six different $\dot \phi_0$. Parameters and initial values except $\dot \phi_0$ are the same with figure \ref{fig:uefftt_1}.
		 (a) Continious (red) line $\dot \phi_0=6.50\,rad\,s^{-1}$, dotted-dashed (green) curve $\dot \phi_0=3.00 \,rad\,s^{-1}$ and dashed (blue) curve $\dot \phi_0=9.00 \,rad\,s^{-1}$.
                 (b) Continious (red) line $\dot \phi_0=69.8\,rad\,s^{-1}$, dotted-dashed (green) curve $\dot \phi_0=67.0 \,rad\,s^{-1}$ and dashed (blue) curve $\dot \phi_0=73.0 \,rad\,s^{-1}$.
                 }
       \label{fig:theta_3}
       \end{center}
\end{figure}

In figure \ref{fig:theta_3}, the change of $\theta$ can be seen for six different $\dot \phi_0$ values when $\theta_0=0.5\,rad$, $\dot \theta_0=0$, $\dot \psi_0=200\,rad\,s^{-1}$, $I_x=0.000228 \,kg \,m^2$, $I_z=0.0000572 \,kg \,m^2$ and $Mgl=0.068J$ which are the same as the case considered in \ref{twoa}.
Different $\dot \phi_0$ values are chosen by considering roots of equation \eqref{quadr}.
It can be seen from figure \ref{fig:theta_3}(a) that when $\dot \phi_0=6.50\, rad\, s^{-1}$, the top precesses regularly (continuous red line); when $\dot \phi_0=3.00\, rad\, s^{-1}$, the top falls or $\theta$ increases (dotted-dashed green curve); and when $\dot \phi_0=9.00\, rad\, s^{-1}$, the top rises or $\theta$ decreases (dashed blue curve). 
It can be seen from figure \ref{fig:theta_3}(b) that when $\dot \phi_0$ is equal to the second root $69.8\, rad\, s^{-1}$, the top precesses regularly (continuous red line);
when $\dot \phi_0$ is a bit smaller than the second root and equal to $67.0\, rad\, s^{-1}$, differently from the previous case the top rises (dotted-dashed blue curve); and when $\dot \phi_0$ is a bit greater than the second root and equal to $73.0\, rad\, s^{-1}$, the top falls (dashed blue curve). 
We shold note that if one of the conditions $\dot \psi_0>0$, $\dot \theta=0$, $\theta_0<\pi/2$ and $I_x>I_z$ were different, then the situation could be different.

After the rise or fall of the top, one may observe one of the different types of motion, e.g. motion with cusps, looping motion or spiraling motion.
We can not determine which kind of motion will be observed only by looking at angular accelerations or initial values. 

\section{Conclusion}
\label{conc}

In the introduction, we have mentioned that some students can wonder about the absence of centripetal or Coriolis forces in the rigid body rotations.
It can be better to give some explanations related to these with a few sentences.
If the rigid body rotates around one of its principal axes, the rigidity of the object provides necessary centripetal forces for each part of the rigid body and total of them simplifies; 
if the rigid body rotates around an axis that is not principal, then the centripetal force must be provided externally.
One can understand the presence of centripetal force by considering the rotation of an elastic body: If an elastic body similar to the top rotates it will elongate which will provide the necessary centripetal force.
For the motion of the heavy symmetric top, rotations in $\theta$ and $\phi$ are not around principal axes and necessary force is provided by the fixed point, otherwise the top slides.
Coriolis force is present when the motion of an object is considered in a rotating reference frame.
Then, there is naturally no Coriolis force in the stationary reference frame, 
and it is not present in the body reference frame since there is not any object moving in the body reference frame.

In section \ref{one}, angular accelerations are calculated in terms of Euler angles from torque-angular momentum relation and Euler equations, 
and the results of both methods are the same.
This explicitly shows that Euler equations are nothing but torque-angular momentum relation written in an accelerated reference frame as it should be.
This equivalence also shows that the extra terms in Euler equations coming from $\vec w \times \vec L$ are inertial torques as expected.

The basic torque-angular momentum relation tells that the rate of change of angular momentum should be equal to torque, and
we have seen that there can be more than one rate of change of angular momentum being equal to the same torque.

In section \ref{twob}, we have seen that any change of angular momentum can cause a sequence of changes in angular velocities and other variables, and all of these changes obey torque-angular momentum relation as expected.
Euler equations or the torque-angular momentum relation are enough to describe rotations of a rigid body,
however, the resultant equations are coupled equations and very hard to understand or explain in that form.
Only some simple relations can be easily understood from these equations, and we have seen examples of this situation in section \ref{two}.

One of the interesting things among the considered cases in section \ref{two} is the relation between the rise of the top and precession angular velocity which requires further attention.
Perry made many different observations on the motion of the symmetric top, and gave an observational principle: "Hurry on the precession, and the body rises in opposition to gravity." \cite{Perry}.
Perry, most probably, carried out his experiments with ordinary fast-spinning tops ($\theta<\pi/2$, $I_x>I_z$ and $Mgl>0$), and for these, there are two $\dot \phi$ values giving the regular precession and both of them have the same sign with $\dot \psi$.
And, during experiments, in general, the regular precession with the smaller $|\dot \phi|$ is observed.
In such cases, if one increases $|\dot \phi|$ a bit, one can observe the rise of the top similar to the situation given in figure \ref{fig:theta_3}(a).
Using this kind of observation, Perry, most probably, gave the mentioned observational principle.
However, the situation is more complex than Perry's observational principle.
This principle is not valid for the greater root, in that case, if one increases $|\dot \phi|$, the top falls.
In addition to this, hurrying precession could fall the top for the regular precession with the smaller root in some special cases, e.g. $\dot \psi<0$, $Mgl<0$, $\theta<\pi/2$ and $I_z>I_x$.
Then, one can say that Perry's observational principle is only valid under certain conditions which are mostly related to fast spinning ordinary symmetric top's motions.

In section \ref{twoc}, we did not consider what will happen after the rise or fall of the top for cases shown in figure \ref{fig:theta_3}.
After the rise or fall, one of the different types of motion can take place which can be found by numerically integrating angular accelerations, i.e. equations \eqref{diffeqnstheta}, \eqref{diffeqnsphi} and \eqref{diffeqnspsi}.
To be able to say what will happen after the rise or fall without numerical integrations, one needs to consider some physical quantities other than angular velocities, angular accelerations and torque.
Indeed, some clues related to these quantities can be found in section \ref{two}: In some cases, we have seen that different components of the angular momentum are conserved, and additionally, we have seen the change in the angular momentum can be understood by considering energy.
We know that $L_z$, $L_{z'}$ and energy are always coserved for the dissipation free cases, which can easily be seen from Lagrangian formalism \cite{Goldstein}. 
From other studies, we know that the usage of conserved angular momenta and energy is necessary to understand the motion of a heavy symmetric top in a better way, and they can be used to determine the motion type \cite{Routh, Tanriverdi_abdffrnt, Tanriverdi_abeql}.
But, this does not mean that Euler equations or torque-angular momentum relation are not enough to describe rigid body rotations.
As mentioned, numerical solutions show that Euler equations and torque-angular momentum relation include all necessary information related to rotations including conservation of angular momenta and energy, however, we are not able to see these by directly looking at these equations.

\end{document}